# The Kagomé-staircase lattice: Magnetic ordering in $Ni_3V_2O_8$ and $Co_3V_2O_8$

N. Rogado[a,*], G. Lawes[b], D. A. Huse[c], A. P. Ramirez[b], and R. J. Cava[a]

[a]*Department of Chemistry and Princeton Materials Institute, Princeton University, Princeton, New Jersey 08544, U. S. A.*
[b]*Los Alamos National Laboratory, Los Alamos, New Mexico 87544, U. S. A.*
[c]*Department of Physics, Princeton University, Princeton, New Jersey 08544, U. S. A.*

**Abstract**

$Ni_3V_2O_8$ and $Co_3V_2O_8$ have spin-1 and spin-3/2 magnetic lattices that are a new anisotropic variant of the Kagomé net, wherein edge-sharing $MO_6$ octahedra form the rises and rungs of a "Kagomé staircase". The anisotropy largely relieves the geometric frustration, but results in rich magnetic behavior. Characterization of the magnetization of polycrystalline samples reveals that the compounds are ferrimagnetic in character. Heat capacity measurements show the presence of four magnetic phase transitions below 9 K for $Ni_3V_2O_8$ and two below 11 K for $Co_3V_2O_8$. Comparison to the low temperature heat capacity of isostructural nonmagnetic $Zn_3V_2O_8$ provides an estimate of the magnetic entropy involved with the phase transitions. The results suggest that $Co_3V_2O_8$ may display magnetic transitions below 2 K.

*PACS*: 75.30.Gw; 75.30.Kz; 75.40.Cx; 75.50.–y

*Keywords*: A. Kagomé net; D. Phase transitions

## 1. Introduction

Geometrically frustrated magnets are of great interest as prototypes of systems where competing interactions result in a large number of energetically degenerate ground states. In these materials, frustration of long range magnetic ordering to temperatures well below the pairwise spin-spin interaction energy is found [1–3]. Geometrical magnetic frustration often results when the arrangement of the component spins is in lattice geometries based on triangles. The two-dimensional Kagomé lattice, consisting of corner-sharing triangles of spins with antiferromagnetic coupling between nearest neighbors, is one of the best known examples of this geometry. This lattice forms the basis for magnetically frustrated layered materials such as $SrGa_4Cr_8O_{19}$ (SCGO) and $BaSn_2ZnGa_4Cr_6O_{22}$ [4–7] and three-dimensionally frustrated materials such as the spinels and the pyrochlores [8–12]. In order to display the large degeneracy of energetically equivalent ground states necessary for strong magnetic frustration, the magnetic lattice must be of high symmetry, with isotropic interactions between spins.

Here we present the synthesis and initial characterization of the magnetic properties of the compounds $Ni_3V_2O_8$ and $Co_3V_2O_8$. The magnetic lattice is based on a new anisotropic variation of the Kagomé net, wherein the pattern of edge-shared transition metal oxide octahedra buckles the Kagome plane into a staircase geometry (Fig.1). The magnetic lattice in this structure type can be occupied by different transition metals [13–16] allowing the study of the spin dependence of the magnetic interactions and ordering schemes. For the present compounds, high-spin $d^8$ $Ni^{2+}$ (spin-1) and $d^7$ $Co^{2+}$ (spin-3/2) occupy the magnetic sites.

The Kagomé staircase layers are separated by nonmagnetic $V^{5+}O_4$ tetrahedra. From the relatively large ratio of interlayer to intralayer metal-metal separation (~1.9), and the more indirect interlayer M-O-O-M superexchange pathway, we expect this structure type to display two-dimensional (2D) magnetic character, with the magnetism dominated by the coupling interactions within the Kagomé magnetic planes.

$Ni_3V_2O_8$ and $Co_3V_2O_8$ have orthorhombic crystal structures [13] characterized by Kagomé-like layers of edge-sharing $Ni^{2+}O_6$ or $Co^{2+}O_6$ octahedra. In both compounds, the $V^{5+}$ separating the $Ni^{2+}$ and $Co^{2+}$ layers is $d^0$ and nonmagnetic. Unlike previously studied Kagomé lattice-based materials, which have flat magnetic layers, the magnetic layers in $M_3V_2O_8$ are buckled, resulting in a Kagomé-staircase geometry. The Kagome staircase planes are perpendicular to the crystallographic *b*-axis. Unlike the case for the ideal

---

[*] Corresponding author. Fax: +1-609-258-6746
    *Email address*: nsrogado@princeton.edu (N. Rogado)

Kagome net, there are two crystallographically inequivalent magnetic sites in the plane of this structure type: the M1 sites, which form the apices of M2-M1-M2 isosceles (though nearly equilateral) triangles, and the M2 sites, which form the bases of the triangles (Fig. 1b). The superexchange interaction between M1 and M2 sites (the "rise" in the staircase) is shown in the figure as $J_1$, while the interaction between two neighboring M2 sites (the "run" in the staircase) is indicated as $J_2$ and runs along the crystallographic $a$-axis. The local geometry around the Ni atoms in $Ni_3V_2O_8$ is shown in Fig. 2 to illustrate the different metal-metal distances and $M^{2+}$-O-$M^{2+}$ bond angles in the magnetic plane. All the Ni atoms are coordinated by 6 oxygen atoms, in an octahedral geometry, and share two oxygen atoms (an octahedron "edge") with each neighboring Ni atom. The Ni2-Ni2 and Ni1-Ni2 separations are quite similar, at 2.97 Å and 2.94 Å, respectively. The $Ni^{2+}$-O-$Ni^{2+}$ bond angles, which mediate the superexchange, are only slightly different from the 90° angle expected for ideal edge-sharing octahedra. The angles in $Ni_3V_2O_8$ are 90.3° and 91.5° for $J_1$ (Ni1-Ni2), and 90.4° and 95.0° for $J_2$ (Ni2-Ni2). The relatively subtle differences in Ni separations and Ni-O-Ni bond angles suggests that $J_1$ and $J_2$ are similar. For $Co_3V_2O_8$, the Co2-Co2 and Co1-Co2 bond distances are 3.02 Å and 2.99 Å, respectively, and the Co-O-Co angles are 90.0° and 91.4° for $J_1$ (Co1-O-Co2), and 89.7° and 95.9° for $J_2$ (Co2-O-Co2), a slightly more anisotropic magnetic system.

## 2. Experimental

Powder samples of $Ni_3V_2O_8$ and $Co_3V_2O_8$ were made using NiO, $Co_3O_4$, and $V_2O_5$ as starting materials. The reagents were mixed thoroughly and heated in dense $Al_2O_3$ crucibles. To make $Ni_3V_2O_8$, the reaction mixture was heated in air at 800 °C for 16 hours. It was then ground, pressed into a pellet, and sintered in an alumina boat at 900 °C for 16 hours. Synthesis of $Co_3V_2O_8$ was carried out by heating the powder mixture at 800 °C under $N_2$ flow for 12 hours, and then at 900 ºC for 12 hours, with intermediate grindings. The resulting powder sample was pressed into a pellet and sintered under $N_2$ flow at 1000 °C for 12 hours, and then at 1100 ºC for 12 hours. Nonmagnetic, isostructural $Zn_3V_2O_8$ was also synthesized, to allow the subtraction of the nonmagnetic contribution to the low temperature heat capacity data of $Ni_3V_2O_8$ and $Co_3V_2O_8$. An intimate mixture of ZnO, $V_2O_5$, and $ZnCl_2$ in the ratio 3:1:2 ($ZnCl_2$ acts as a flux) was heated at 600 °C for 4 hours in air and furnace-cooled to room temperature.

Powder X-ray diffraction (XRD) with Cu Kα radiation was employed to characterize the samples. DC magnetization measurements were obtained using a Quantum Design PPMS magnetometer. Heat capacity measurements were performed on sintered powder samples using a standard semi-adiabatic heat pulse technique at zero applied field.

## 3. Results and discussion

The samples were found to be single phase by powder X-ray diffraction. The crystal structures were found to be orthorhombic (space group *Cmca*) with $a$ = 5.931(3) Å, $b$ = 11.374(3) Å, and $c$ = 8.235(3) Å for $Ni_3V_2O_8$, and with $a$ = 6.045(4) Å, $b$ = 11.517(3) Å, and $c$ = 8.316(3) Å for $Co_3V_2O_8$ (tungsten used as an internal standard). These values are in close agreement with previously reported data [13].

The magnetic susceptibilities, $\chi(T)$, for $Ni_3V_2O_8$ and $Co_3V_2O_8$ in an applied field of 1 Tesla are shown in Fig. 3. $\chi(T)$ of $Ni_3V_2O_8$ (upper panel), where all the magnetic sites are filled with S = 1 spins, shows a broad maximum at around 3.7 K that is suggestive of antiferromagnetic (AFM) ordering. In the case of $Co_3V_2O_8$ (lower panel), which has S = 3/2 spins occupying the magnetic lattice, $\chi(T)$ shows a ferromagnetic (FM) transition at around 6 K. The inverse susceptibility $1/\chi$ plots for $Ni_3V_2O_8$ and $Co_3V_2O_8$ over a broader temperature range are shown in the figure as insets. The high temperature data, from 250-300 K, was fitted to the Curie-Weiss law from which the Curie ($C$) and Weiss ($\theta$) constants were determined. There is a slight curvature in the high-temperature $\chi(T)$ data up to 300 K. Thus, the parameters obtained from the present Curie-Weiss fits should only be taken as an approximation. The Curie-Weiss fit to the $1/\chi$ data of $Ni_3V_2O_8$ at 1 Tesla yielded $C$ = 1.37 emu K/ mol Ni and $\theta$ = −30 K. The negative $\theta$ value implies that AFM interactions dominate between the $Ni^{2+}$ spins. In the case of $Co_3V_2O_8$, the Curie-Weiss fit yielded $C$ = 3.29 emu K/ mol Co and a positive $\theta$ value ($\theta$ = +14 K), indicative of dominant FM coupling interactions between the $Co^{2+}$ spins. The effective magnetic moments ($\mu_{eff}$) for Ni or Co were calculated to be 3.3 $\mu_B$ and 5.1 $\mu_B$, respectively. The experimentally obtained magnetic moments are noticeably larger than the simple prediction of a spin-only system (2.83 $\mu_B$ for S = 1 and 3.87 $\mu_B$ for S = 3/2). However, the $\mu_{eff}$ values obtained for both compounds are still within the range of experimental values typically observed for $Ni^{2+}$ and $Co^{2+}$ in high spin complexes [17], suggesting the presence of an orbital contribution to the overall magnetic moment.

The field dependence of the magnetization of $Ni_3V_2O_8$ and $Co_3V_2O_8$ is shown in Fig. 4. The M vs. H plot for $Ni_3V_2O_8$ (upper panel) taken at 2 K shows a magnetic hysteresis. On the other hand, field-dependent magnetization measurements for $Co_3V_2O_8$ (lower panel) taken at 5 K shows no hysteresis at all but only weak ferromagnetism. In the M/H plot for $Co_3V_2O_8$ at 10 K, two inflections are observed: one at H = 0.3 Tesla and another at H = 1 Tesla. This may be indicative of the presence of a potentially interesting temperature-field magnetic phase diagram.

The low temperature heat capacity raw data for $Ni_3V_2O_8$ and $Co_3V_2O_8$ are shown in the upper panel of Fig. 5. The heat capacity of $Ni_3V_2O_8$ (upper panel) clearly shows four distinct transitions: at 2.6, 4, 6.4, and 9 K. No indication of the transitions at 6 and 9 K are observed in $\chi(T)$ (lower panel, inset), even with high resolution measurements, though there are changes observed between 2.6 and 4 K for $\chi(T)$ taken at different fields. The magnetic contribution to the heat capacities $C_m$ was obtained by subtracting the lattice contribution from the heat capacity raw data. The lattice contribution was estimated from the heat capacity data of $Zn_3V_2O_8$ (also shown in the figure, upper panel), the isostructural nonmagnetic analog. Integration of $C_m$/T against T up to 30 K (assuming a smooth curve from 2 to 0 K) gives an entropy of ~7.6 J/mol Ni·K. This value is approximately equal to the total entropy needed to fully disorder one mole of a spin-1 system. In the case of $Co_3V_2O_8$, two magnetic transitions can be seen in the heat capacity (upper panel): a small peak at 11 K, and a large, sharp $\lambda$-type anomaly at 6 K, both of which are observable in the $\chi(T)$ data (lower panel). The integrated magnetic entropy in this material up to 30 K can be estimated as ~5 J/mol Co·K. This is less than half the value expected for a spin-3/2 system, indicating that only a fraction of the spins are participating in the observed ordering transitions. It would be of interest to explore the possible existence of magnetic transitions in $Co_3V_2O_8$ below T = 2 K, which is the limit of our measurements.

## 4. Conclusion

The magnetic properties of polycrystalline samples of S = 1 and S = 3/2 Kagomé lattice based $Ni_3V_2O_8$ and $Co_3V_2O_8$ are reported. Unlike previously studied planar Kagomé lattice-based materials, the magnetic layers in these compounds are buckled, resulting in a "staircase" geometry. The lower symmetry, i.e. the buckled Kagomé layers and the resulting inequivalent superexchange interactions within the magnetic lattice, contribute to the reduction of the effects of geometric frustration in these compounds. More importantly, the anisotropic magnetic coupling in the stair-like Kagomé layers leads to the appearance of distinct multiple temperature- and field-dependent magnetic phase transitions. The data obtained on polycrystalline samples gives a general indication of the complexity of magnetic ground states in these examples of the spin-1 and spin-3/2 anisotropic Kagomé lattice. The study of single crystals to obtain more detailed information will be of significant interest.

## Acknowledgments

This research was supported by the National Science Foundation through the MRSEC program (NSF MRSEC grant DMR-9809483).

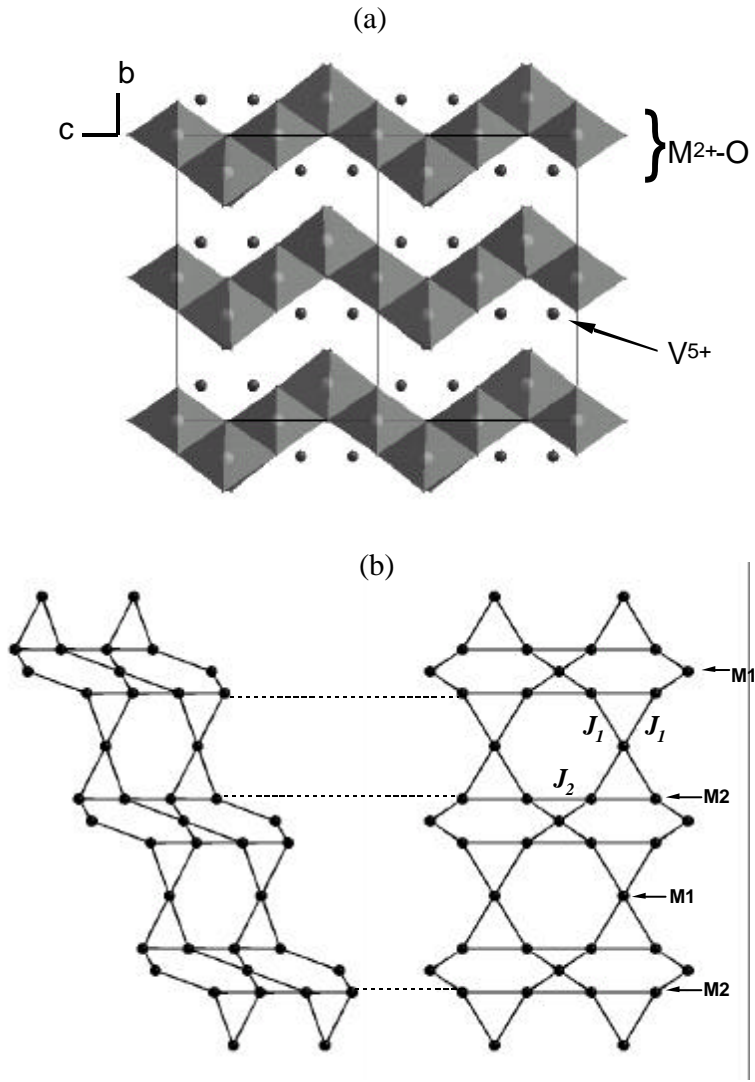

Fig. 1. (a) The crystal structure of $M_3V_2O_8$ showing layers of edge-sharing $MO_6$ octahedra separated by $V^{5+}$ ions. (b) Schematic representation of the "Kagomé-staircase" lattice found in $M_3V_2O_8$ showing the magnetic atoms only.

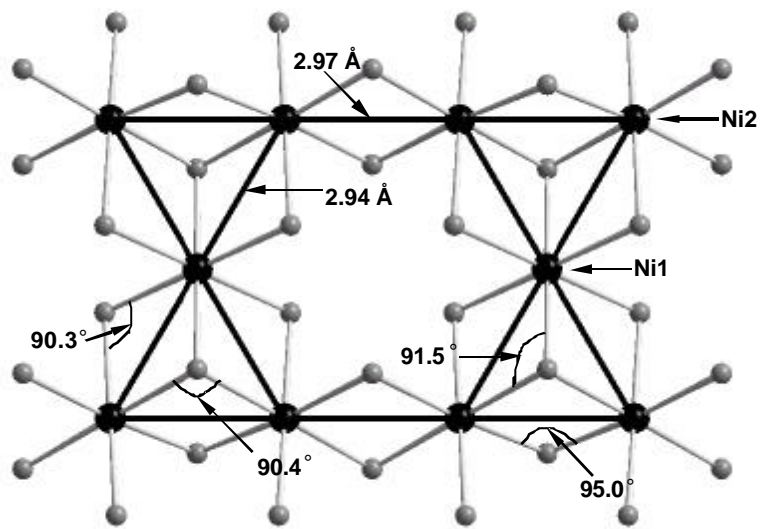

Fig. 2. Local geometry around the Ni atoms in Ni$_3$V$_2$O$_8$ indicating the different Ni-Ni separations and Ni-O-Ni bond angles. Black circles are Ni atoms, gray circles are oxygen atoms.

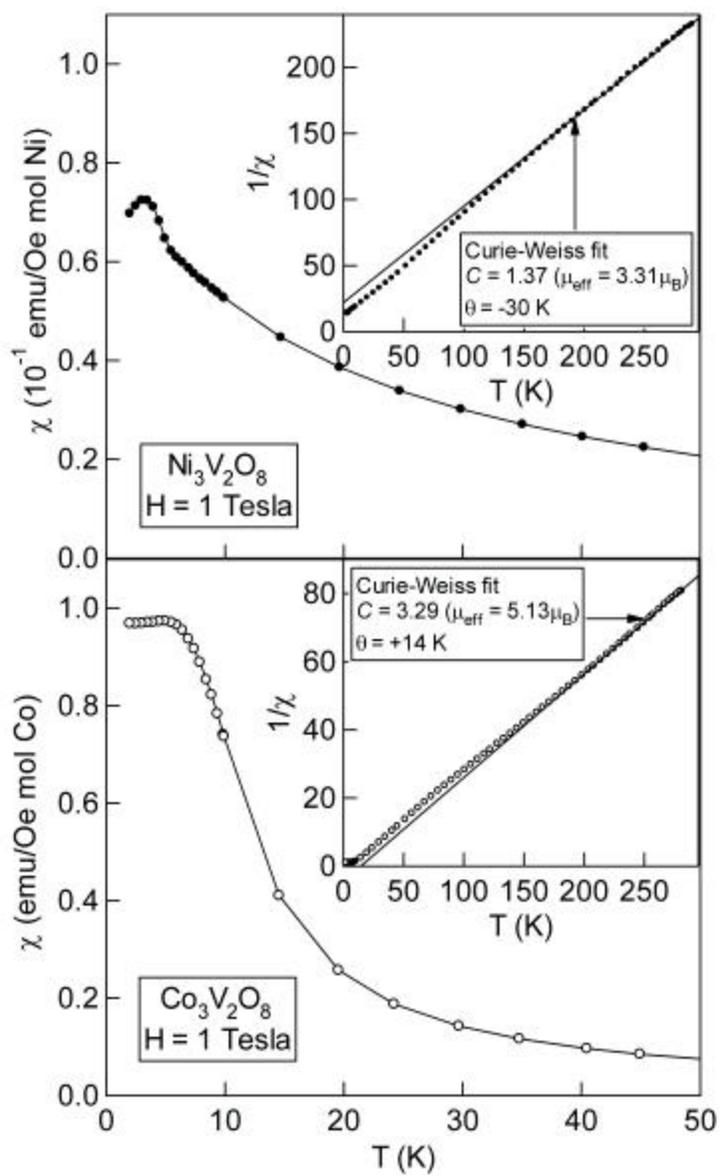

Fig. 3. Temperature dependence of the magnetic susceptibilities of $Ni_3V_2O_8$ (upper panel) and $Co_3V_2O_8$ (lower panel) in an applied field of 1 Tesla. Insets show the corresponding inverse susceptibility plots.

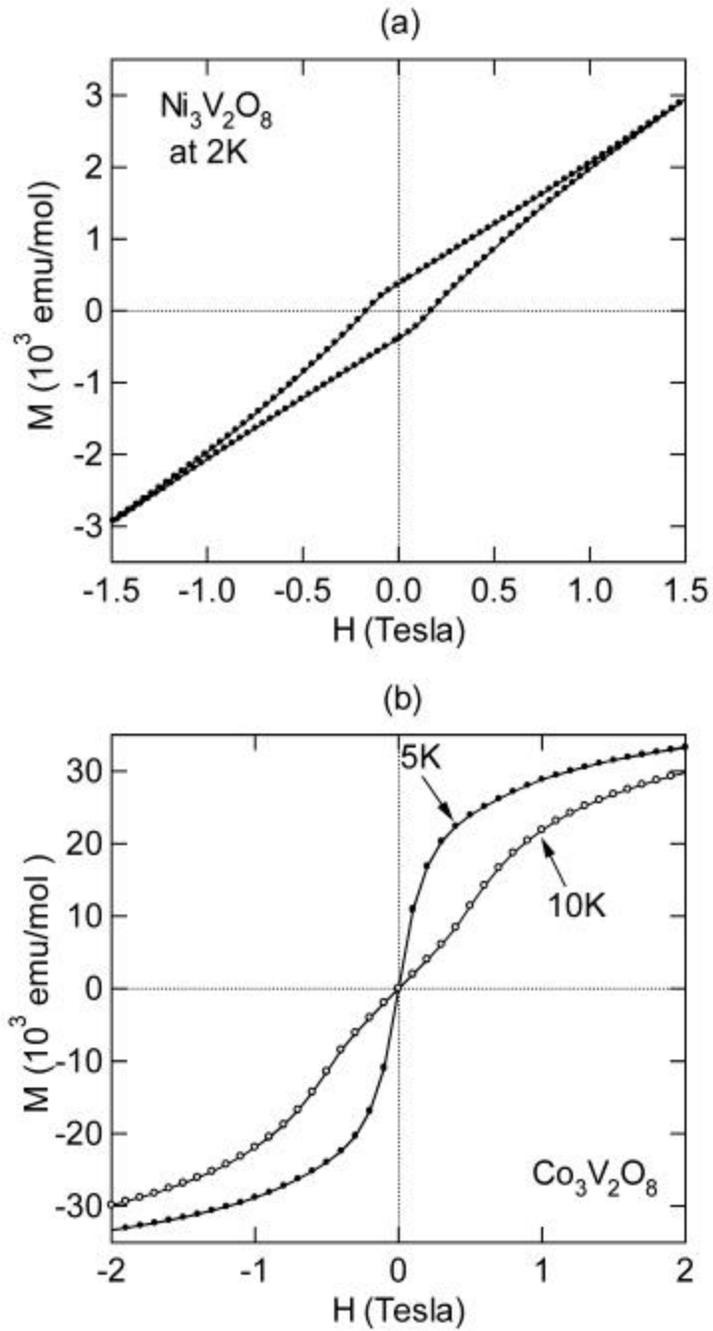

Fig. 4. Magnetization vs. applied field at T = 2K for $Ni_3V_2O_8$ (upper panel), and at T= 5K and 10K for $Co_3V_2O_8$ (lower panel).

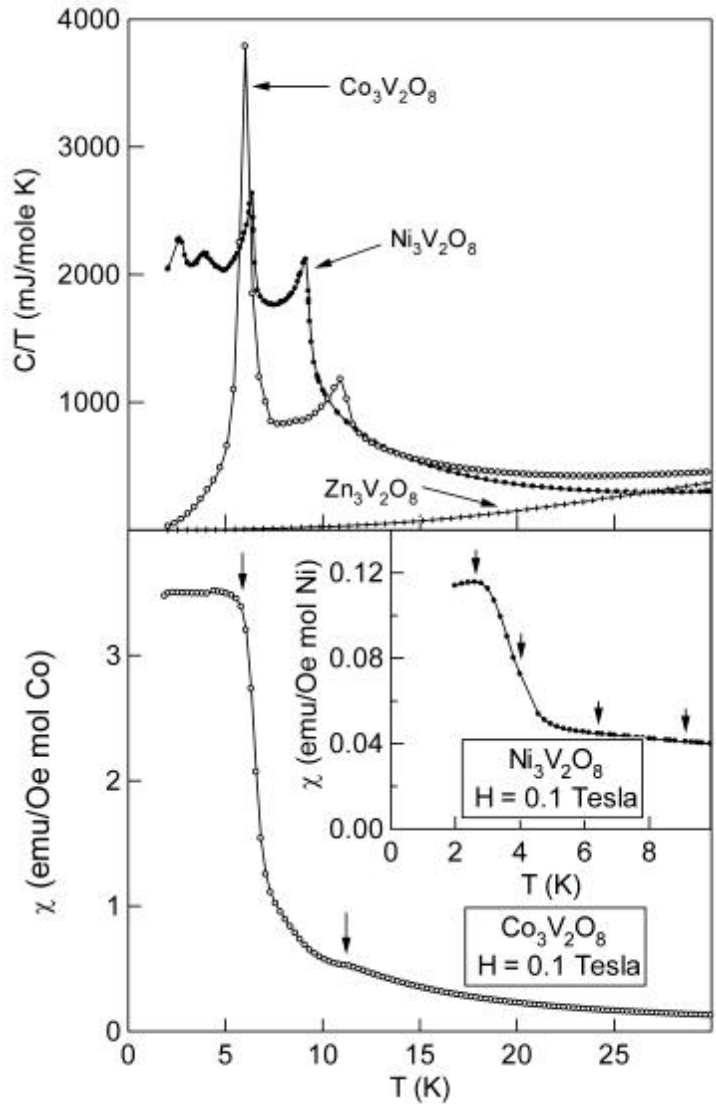

Fig. 5. Upper panel: Temperature dependence of the heat capacities of $Ni_3V_2O_8$ and $Co_3V_2O_8$ represented as $C/T$ vs. T. The heat capacity of nonmagnetic isostructural $Zn_3V_2O_8$ is also plotted. Lower panel: High-resolution measurements of the susceptibilities of $Co_3V_2O_8$ and $Ni_3V_2O_8$ (inset) at 0.1 Tesla. Arrows pointing downwards indicate where the magnetic transitions are observed in the corresponding heat capacity data.

**References**

[1]  A. P. Ramirez, Annu. Rev. Mater. Sci. 24 (1994) 453.
[2]  J. E. Greedan, J. Mater. Chem. 11 (2001) 37.
[3]  R. Moessner, Can. J. Phys. 79 (2001) 1283.
[4]  X. Obradors, A. Labarta, A. Isalgué, J. Tejada, J. Rodriguez, and M. Pernet, Solid State Commun. 65 (1988) 189.
[5]  A. P. Ramirez, G. P. Espinosa, and A. S. Cooper, Phys. Rev. B 45 (1992) 2505.
[6]  A. P. Ramirez, G. P. Espinosa, and A. S. Cooper, Phys. Rev. Lett. 64 (1990) 2070.
[7]  I. S. Hagemann, Q. Huang, X. P. A. Giao, A. P. Ramirez, and R. J. Cava, Phys. Rev. Lett. 86 (2001) 894.
[8]  A. P. Ramirez, A. Hayashi, R. J. Cava, R. Siddharthan, and B. S. Shastry, Nature 399 (1999) 333.
[9]  S. T. Bramwell and M. J. P. Gingras, Science 294 (2001) 1495.
[10]  S.-H. Lee, C. Broholm, W. Ratcliff, G. Gasparovic, Q. Huang, T. H. Kim, and S.-W. Cheong, Nature 418 (2002) 856.
[11]  J. E. Greedan, C. R. Wiebe, A. S. Wills, and J. R. Stewart, Phys. Rev. B 65 (2002) 184424.
[12]  S.-H. Lee, Y. Qiu, C. Broholm, Y. Ueda, and J. J. Rush, Phys. Rev. Lett. 86 (2001) 5554.
[13]  E. E. Sauerbrei, R. Faggiani, and C. Calvo, Acta Cryst. B 29 (1973) 2304.
[14]  H. Fuess, E. F. Bertaut, R. Pauthenet, and A. Durif, Acta. Cryst. B 26 (1970) 2036.
[15]  R. D. Shannon and C. Calvo, Can. J. Chem. 50 (1972) 3944.
[16]  R. Gopal and C. Calvo, Can. J. Chem. 49 (1971) 3056.
[17]  K. Burger, Coordination Chemistry: Experimental Methods, London, 1973.